\documentstyle[psfig,times]{mn}

\def\spose#1{\hbox to 0pt{#1\hss}}
\def\lta{\mathrel{\spose{\lower 3pt\hbox{$\mathchar"218$}}
     \raise 2.0pt\hbox{$\mathchar"13C$}}}
\def\gta{\mathrel{\spose{\lower 3pt\hbox{$\mathchar"218$}}
     \raise 2.0pt\hbox{$\mathchar"13E$}}}                  
 
\title[A CMB decrement towards a cluster of mJy
radiosources]{Detection of a CMB decrement towards a cluster of
mJy radiosources}
 
\author[Garret~Cotter et al.]
{
Garret Cotter,$^{1}$
Helen J. Buttery,$^{1}$
Steve Rawlings,$^{2}$
Steve Croft,$^{2}$
Gary J. Hill,$^{3}$
Pamela Gay,$^{3}$\cr
Rhiju Das,$^{1}$\thanks{Present address: Physics Department, Stanford
University, CA 94305-4060, USA}
Niv Drory,$^{4}$
Keith Grainge,$^{1}$
William F. Grainger,$^{1}$\cr
Michael E. Jones,$^{1}$
G. G. Pooley,$^{1}$
Richard Saunders$^{1}$\\
$^{1}$Astrophysics, Cavendish Laboratory, Madingley Road, Cambridge CB3 0HE, UK\\
$^{2}$Astrophysics, Department of Physics, Keble Road, Oxford OX1 3RH,
UK\\
$^{3}$University of Texas at Austin,  Astonomy Department, RLM
15.308, TX 78712-1083, USA\\
$^{4}$Universit\"{a}ts-Sternwarte M\"{u}nchen,  Scheinerstrasse 1, D-81679 M\"unchen, Germany
}

\date{Submitted 2001 March 1; revised 2001 August 22; accepted 2001
September 25}

\begin{document}

\maketitle
 
\begin{abstract}

We present the results of radio, optical and near-infrared
observations of the field of TOC~J0233.3+3021, a cluster of
milliJansky radiosources from the TexOx Cluster survey. In an
observation of this field with the Ryle Telescope (RT) at 15 GHz, we
measure a decrement in the cosmic microwave background (CMB) of $-675
\pm 95 \mu$Jy on the RT's  $\approx$ 0.65 k$\lambda$ baseline. Using 
optical and infrared imaging with the McDonald 2.7-m Smith Reflector,
Calar Alto 3.5-m telescope and UKIRT, we identify the host galaxies of
five of the radiosources and measure magnitudes of $R
\approx 24$, $J \approx 20$, $K \approx 18$.

The CMB decrement is consistent with the Sunyaev-Zel'dovich (SZ)
effect of a massive cluster of galaxies, which if modelled as a
spherical King profile of core radius $\theta_C = 20^{\prime\prime}$
has a central temperature decrement of $900~\mu$K. The magnitudes and
colours of the galaxies are consistent with those of old ellipticals
at $z \sim 1$. We therefore conclude that TOC~J0233.3+3021 is a
massive, high redshift cluster. These observations add to the growing
evidence for a significant population of massive clusters at high
redshift, and demonstrate the effectiveness of combining searches for
AGN `signposts' to clusters with the redshift-independence of the SZ
effect.

\end{abstract} 


\section{Introduction} 

Given a set of cosmological parameters such as $\Omega_M$ and
$\Omega_{\Lambda}$, $N$--body simulations of the build--up of the mass
function of dark matter halos (see e.g. Jenkins et al, 2000) are now
sufficiently advanced that there are clear predictions concerning the
number of systems of a given mass expected at a given cosmic epoch. At
redshifts $z \approx 1$ and beyond, however, only a few members of the
population of galaxy clusters are known (e.g. Stanford et al. 1997;
Rosati et al. 1999; Chapman et al. 2000; Blanton et al. 2000; Fabian
et al. 2001). Here we present the first result of a new technique
designed to find distant clusters.

The TexOx Cluster (TOC) survey is an initiative designed to find
massive high-redshift clusters. The TOC survey searches for
statistical overdensities of radiosources in the deep, wide area NRAO
VLA Sky Survey (NVSS; Condon et al. 1998). Although any search for
clusters using AGNs as signposts will find only a small fraction of
the total population, the surface density of radiosources is
sufficiently low, and the sky coverage of the NVSS sufficiently large,
that this technique is very efficient. TOC selects candidates where
the radiosource overdensity is greater than 5-$\sigma$ in a $7 \times
7$ arcmin$^2$ box. High-resolution VLA followup is used to eliminate
the small number of cases where the apparent clustering of sources is
spurious, and then optical and infrared imaging are carried out to
identify the hosts galaxies of the radiosources and other candidate
clusters galaxies.

A full description of the first TOC results will be presented in a forthcoming
paper (Croft et al. in preparation; hereafter C2001). In this letter we
describe observations of one of the TOC candidates, TOC~J0233.3+3021. Our
particular motivation in these observations was to exploit the
Sunyaev-Zel'dovich (SZ, Sunyaev and Zel'dovich 1972) effect as a means of
confirming candidate high-redshift clusters.

In any cluster of galaxies, electrons in the intracluster plasma
scatter the CMB photons passing through, distorting the thermal
spectrum of the incident CMB and producing a change in brightness
temperature given by $ \Delta T_{\rm SZ} \propto \int n_eT_e \, {\rm
d}l$. Here $n_e$ and $T_e$ are the electron number density and
Temperature, and d$l$ is an element of line-of-sight element through
the cluster. Two consequences of this are:

\begin{itemize}

\item{Because the CMB imprinting process is a scattering, the same
$\Delta T_{SZ}$ is produced by a given $\int n_e T_e \, {\rm d}l$ whatever the
redshift of the cluster.}

\item{The SZ signal from a cluster is less sensitive to mass
concentration than either X-ray (emissivity $\propto \int n_e ^2 \, {\rm
d}l$) or optical searches. The integral of $\Delta T_{SZ}$ over the
projected surface of a cluster is proportional to just the
mass--temperature product of the gas.}

\end{itemize}

Observations of the SZ effect in several X-ray selected clusters at $z
\gta 0.5$ have now been made using cm-wave interferometers (e.g. Grego
et. al 2001, Joy et al. 2001, Grainge et al. 2001). Indeed, the SZ effect is an
ideal tool for conducting a survey for clusters, not only because it allows
cluster selection to high redshift, but because the selection is limited only
by gas mass, almost independently of redshift. Plans have been developed for a
new generation of dedicated SZ survey telescopes (e.g. Kneissl et al. 2001,
Holder et al. 2000); however, in the absence of an SZ survey, we have
undertaken pointed SZ observations of a TOC candidate to confirm that it is a
high-$z$ cluster and estimate its mass.

\section{Observations and results}

\subsection{Ryle Telescope Observations}

The standard configuration of the Ryle Telescope (RT) for SZ
observations has described in detail elsewhere (e.g. Grainge et
al. 1996). A nearly-east-west array of five 13-m antennas is used,
operating at 15~GHz, giving projected baselines from 13m to 108m
(0.65--5.4 k$\lambda$). Unresolved sources in the telescope field of
view --- typically radiogalaxies --- contribute equal flux to all
baselines of the array, whereas the extended structure of any SZ
signal is typically present only on the shortest baseline. We
therefore use the long baselines to identify and remove point sources
from the field, and use the residual flux on the shortest baseline to
measure the SZ flux decrement.

\begin{figure}
\begin{center}
\psfig{figure=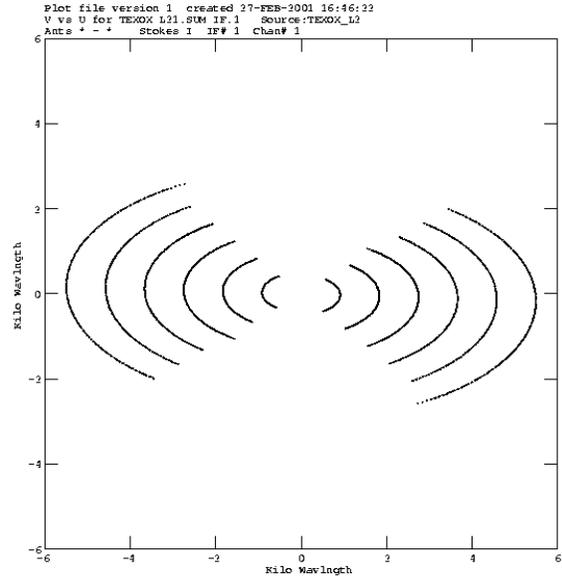,width=\linewidth,clip=}
\caption{Aperture-plane coverage for the RT observations of the TOC~J0233.3+3021
field. Two physical baselines contribute to the shortest baseline in
this figure, three to the second-shortest, two to the third shortest,
and one to each of the three longest baselines.}
\label{fig:uv_plane}
\end{center}
\end{figure}

RT observations of the TOC~J0233.3+3021 field totalling 280 hours
were made during 35 days in the period 1999 January to 1999 June (the
declination of the source means that it can only be observed for 8
hours each day).  The telescope was in Cb configuration (Grainge et
al. 1996), resulting in the aperture-plane coverage shown in Fig.\
\ref{fig:uv_plane}. In this configuration the array provides two baselines
of maximum projected length 900$\lambda$, three of length 1800$\lambda$, two of 
2700$\lambda$ and one each of 3600$\lambda$, 4500$\lambda$ and 5400$\lambda$.
Note that since each day's observation is only eight hours, the
baseline tracks in the $uv$ plane are azimuthally truncated. For
each day, observations of the target field were interleaved with
observations of a phase calibrator about every 20 minutes, and a
primary flux calibrator (3C286 or 3C48) was observed at either the
start or end of the run. 

\begin{figure}
\begin{center}
\psfig{figure=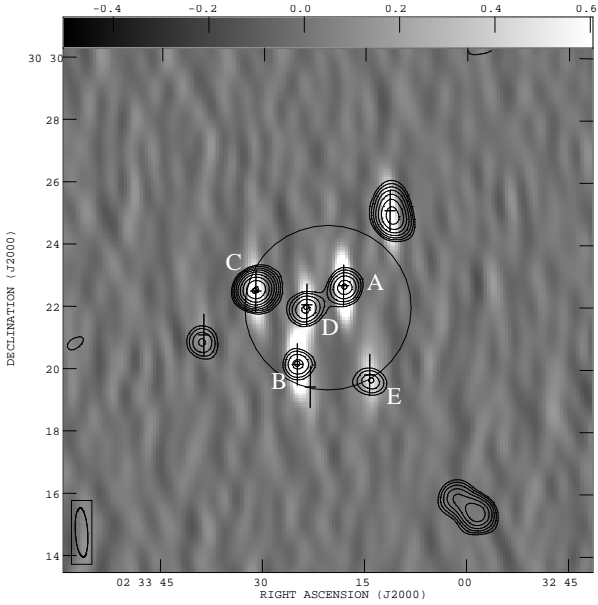,width=\linewidth,clip=}
\caption{{\sc Clean}ed 1.5--5.4k$\lambda$ Ryle Telescope map of the TOC~J0233.3+3021
field before source subtraction  overlaid with the NVSS 1.4-GHz
map of the field (contours). The greyscale runs from $- 0.5$ to
0.6~mJy~beam$^{-1}$, and the contours are 3, 3$\sqrt 2$, 6, 6$\sqrt 2$,
\dots multiples of 0.6 mJy~beam$^{-1}$. The large circle shows the
FWHM of the RT primary beam, and the ellipse at lower left shows the FWHM of
the RT synthesized beam. The five NVSS sources which define the TOC survey
overdensity are labelled A---E; the crosses show the positions of the point
sources found at 15 GHz. Details of these sources are listed in Table
\ref{table:RT_sources_opt_ir_mags}.}
\label{fig:nvss_plus_long_baselines}
\end{center}
\end{figure}

\begin{figure}
\begin{center}
\psfig{figure=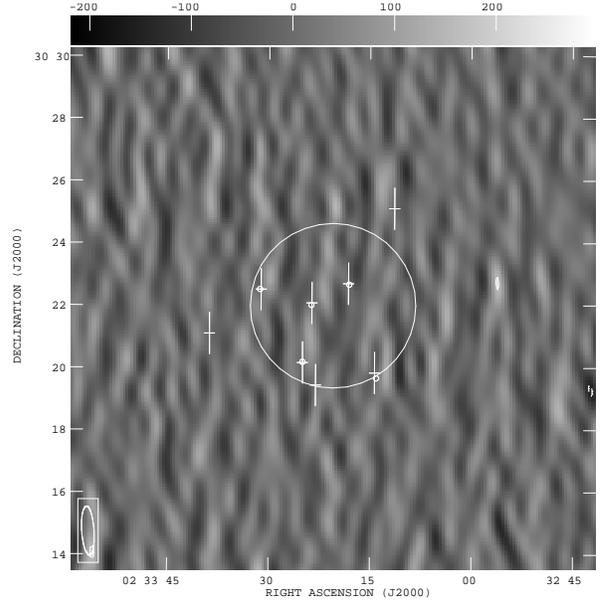,width=\linewidth,clip=}
\caption{1.5--5.4k$\lambda$ Ryle Telescope map of the TOC~J0233.3+3021
field after subtraction of the point sources listed in Table
\ref{table:RT_sources_opt_ir_mags}. The greyscale runs from $- 0.217$ to
0.296 mJy~beam$^{-1}$, and contours are plotted at -3 and 3 times 0.050
mJy~beam$^{-1}$, the map rms outside the primary beam. The map is
consistent with noise. The primary and synthesized beams are shown as in
Fig.\ \ref{fig:nvss_plus_long_baselines}. The small circles are positions of
the five radiogalaxy optical IDs given in Table 1.}
\label{fig:long_baselines_source_subtracted}
\end{center}
\end{figure}

We analysed the combined 35-day visibility data for evidence of a flux
decrement on the shortest baseline. First, we made and {\sc Clean}ed a
naturally-weighted map, using only visibilities from projected
baselines longer than 1.5~k$\lambda$, which is shown in Fig\
\ref{fig:nvss_plus_long_baselines},  with the NVSS image of the
TOC~J0233.3+3021 field overlaid. Several of the NVSS sources are
readily identifiable on the RT long-baseline map. Next, two of us (HJB
and GC) independently investigated the sources present in the RT
long-baseline map. We measured the positions of sources in the map
using the task {\sc Maxfit} in {\sc Aips}, then subtracted the
brightest source from the visibility data using {\sc Uvsub} in {\sc
Aips}. This process was repeated until no sources remained whose peak
fluxes were greater than 3.5 times the rms in the map well outside the
primary beam. A list of eight sources was agreed upon; all but one of
these sources is identifiable on the NVSS map, as shown in Fig\
\ref{fig:nvss_plus_long_baselines}. Accurate positions and fluxes for
these sources were then calculated using the the {\sc FluxFitter}
algorithm (Grainger et al. 2001). {\sc FluxFitter} takes input
estimates of the point-source fluxes and positions, and varies these
to minimize $\chi$-squared between the visibility data and model
visibilities containing the point sources and noise; it has been
demonstrated to produce robust flux and position measurements
(Grainger et al. 2001). The resulting source positions and fluxes,
calculated using all the data from the projected baselines longer than
1.5~k$\lambda$ shown in Table\
\ref{table:RT_sources_opt_ir_mags}.  These sources were then
subtracted
\begin{table*}

\label{table:RT_sources_opt_ir_mags}

\caption{Details of the radiosources in the TOC~J0233.3+3021
field, in the following format: Position of source at 15-GHz as
subtracted from RT observations; flux at 15 GHz; Position of host
galaxy identification (C2001) ; $R$-, $J$- and $K$-band magnitudes
(C2001); Label used in  Fig. 2.}

\begin{tabular}{|l|l|l|l|l|l|l|l|l|}
\hline

RT RA (J2000)  &  RT Dec (J2000)  & $S_{\rm 15 GHz}$   &  ID RA (J2000) & ID Dec (J2000) & $R$   & $J$   & $K$    &  Label     \\
               &                  &    $/$  mJy        &                &                &       &       &        &            \\     
\hline						       							   	    
						       							   	    
02 33 18.22    &  30 22 42.91     &    2.007           &    02 33 18.1  & 30 22 40.1     & 23.75 & 19.75 & 18.19  & A         \\
02 33 24.97    &  30 20 09.31     &    1.906           &    02 33 25.0  & 30 20 11.2     & 23.28 & 19.74 & $>$ 19 & B         \\
02 33 11.41    &  30 25 07.91     &    1.118  	       &                &                &       &       &        &           \\
02 33 31.17    &  30 22 33.31     &    1.077           &    02 33 31.3  & 30 22 31.2     & 23.35 & 20.76 & 18.58  & C         \\
02 33 23.65    &  30 22 06.41     &    1.053           &    02 33 23.7  & 30 22 00.9     & 23.10 & 19.86 & 18.19  & D         \\
02 33 14.23    &  30 19 46.41     &    0.497           &    02 33 14.1  & 30 19 39.8     & 24.96 & 20.82 & 18.39  & E         \\
02 33 38.87    &  30 21 10.41     &    0.352   	       &                &                &       &       &        &           \\ 
02 33 22.83    &  30 19 24.11     &    0.339   	       &                &                &       &       &        &           \\ 

\hline
\end{tabular}

\end{table*}
from all the visibilites.  To check that no residual flux from any of
the sources was present, source-subtracted maps were made using
baselines of length 1.5~k$\lambda$ and greater (shown in Fig\
\ref{fig:long_baselines_source_subtracted}), and 2.0~k$\lambda$ and
greater. These maps are both consistent with noise. 

\begin{figure}
\begin{center}
\psfig{figure=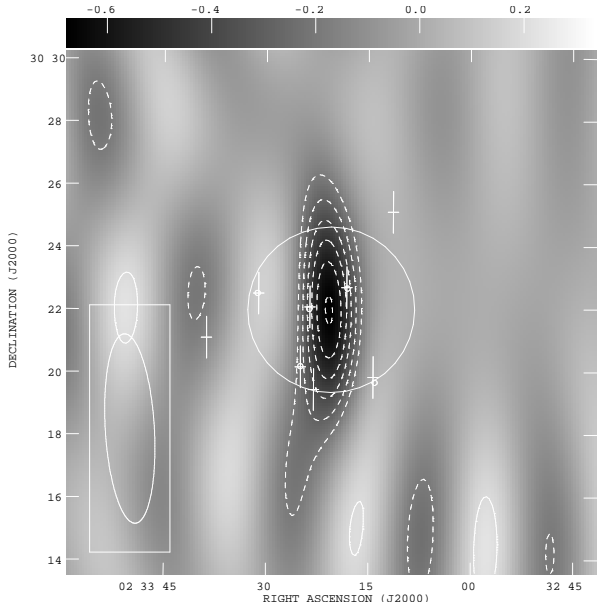,width=\linewidth,clip=}
\caption{{\sc Clean}ed 0.65--1.0 k$\lambda$ Ryle Telescope map of the TOC~J0233.3+3021 
field after subtraction of the point sources listed in Table
\ref{table:RT_sources_opt_ir_mags}. The greyscale runs from $-0.675 $ to
0.335 mJy~beam$^{-1}$. Contours are plotted at -7,-6,-5,-3,-4,-2 and 2
times 0.095 mJy~beam$^{-1}$, whihc is consistent with the map rms well
outside the primary beam, as described in Section 3.2. Beams and
source positions are shown as in previous figures.}
\label{fig:short_baselines_source_subtracted}
\end{center}
\end{figure}

\begin{figure}
\begin{center}
\psfig{figure=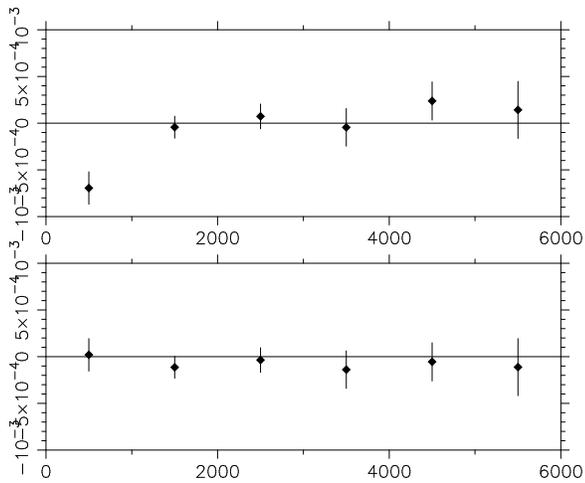,width=\linewidth,angle=-90,clip=}
\caption{Real (top) and imaginary (bottom) parts of the
visibilities, azimuthally averaged about the centre of the decrement after
source subtraction. The baseline scale ranges from 0 to 6000 wavelengths,
and the flux scale in each plot ranges from $-1.0$ to 1.0 mJy. The
visibilities indicate a single extended negative source at the centre of
the field.}
\label{fig:azimuthal_average_plot}
\end{center}
\end{figure}

Finally, a map was made using source-subtracted visibilities with
projected baselines in the range 0.65--1.0~k$\lambda$, corresponding
to the single shortest physical baseline. This map showed a negative
feature centred at 02$^h$ 33$^m$ 20.8$^s$, 30$^\circ$ 21$^\prime$ 58.7$^{\prime \prime}$ (J2000), and with a minimum
flux of in the dirty map of $-675
\mu$Jy. The synthesized beam of this single-baseline map has strong
sidelobes, so the rms level on the map even well outside the primary
beam is still somewhat affected by the presence of the negative
feature at the centre of the map. We therefore made a very large map
(2048 $\times$ 2048 5$^{\prime \prime}$-square pixels) and measured the map rms in
128-pixel square regions where the map apppeared by eye to be
unaffected by sidelobes from the central source. The rms levels agreed
wth the expected noise, due to the telescope system temperature, of
95 $\mu$Jy. The short-baseline map was {\sc Cleaned} to a flux limit
of $- 285
\mu$Jy, and this map is presented in Fig.\
\ref{fig:short_baselines_source_subtracted}. On inspection of the azimuthally-averaged 
visibilities  (Fig.\
\ref{fig:azimuthal_average_plot}), the flux on the shortest
baseline is $-695 \mu$Jy,  with an rms of 170  $ \mu$Jy (note that
since the shortest baseline has an elliptical track in the aperture
plane, this rms includes the variation in the amplitude of the SZ
signal as a function of  $uv$ distance).

\subsection{Optical and infra-red observations of the galaxies}

We obtained $R$-band imaging of the TOC~J0233.3+3021 field using the
Imaging Grism Instrument on the McDonald Observatory 2.7-m Smith
Reflector in 1998 October, 1999 October, and 2000 September, and
$J$-band imaging of the field at the Calar Alto 3.5-m telescope in
1999 August. Additionally, we obtained several $K$-band images of the
TOC~J0233.3+3021 field with UFTI at UKIRT in 2000 July. Full details,
and extensive discussion of other candidate cluster galaxies, will be
presented in C2001.  Using these images, in combination with
high-resolution VLA imaging (C2001), we have identified the host
galaxies of the central five NVSS radiosources shown in
Fig. \ref{fig:nvss_plus_long_baselines}. For four of these galaxies,
we have obtained 3.5$^{\prime
\prime}$-diameter aperture magnitudes in all filters; for the fifth
galaxy, only $R$ and $J$ magnitudes are available.  The measured
magnitudes of the galaxies are presented in Table
\ref{table:RT_sources_opt_ir_mags}\, along with the positions of the host galaxy
identifications. The positions were measured from the $R$-band image,
with an astrometric solution using the positions of stars measured in
the Cambridge APM survey. The magnitudes of the galaxies are
all roughly $K \approx 18$, $J \approx 20$, $R \approx 24$.

In 2000 January, we attempted spectroscopy of the radiogalaxies with
ISIS on the WHT. We used a 2-arcsec slit, with sensitivity on the ISIS
red arm up to 9340 \AA. We took 1800-second spectra of A (at PA=118,
the axis of its double structure; see C2001), galaxies B \& C (as a
pair with PA=113), and galaxies D \& E (as a pair with PA=77). Faint red
continuum was detected spectroscopically in all the cases expected
(i.e. the 4 brightest), but no emission lines were evident.

\section{Discussion}

\subsection{The redshifts of the radiogalaxies}

\begin{figure}
\begin{center}
\psfig{figure=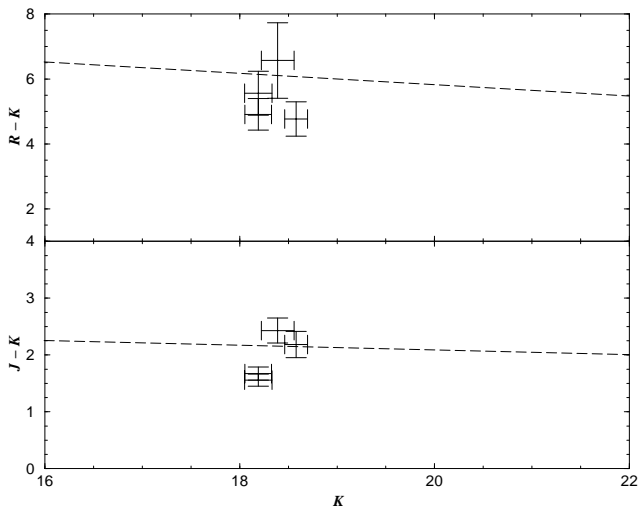,width=\linewidth,clip=,angle=-90}
\caption{$R - K$ and $J - K$ colour-magnitude plots for the host
galaxies of the radiosources; cf Fig. 4 of Stanford et al. (1997). The dashed
lines show the red sequence of the galaxies in the $z = 1.27$ cluster
identified by Stanford et al. (1997). The similar red colours of the
TOC~J0233.3+3021 galaxies suggests that they are old elliptical galaxies, with
the $R$ and $J$ filters straddling the observed-frame 4000-\AA\ break.}
\label{fig:rjk_colmag_plot}
\end{center}
\end{figure}

\begin{figure}
\begin{center}
\psfig{figure=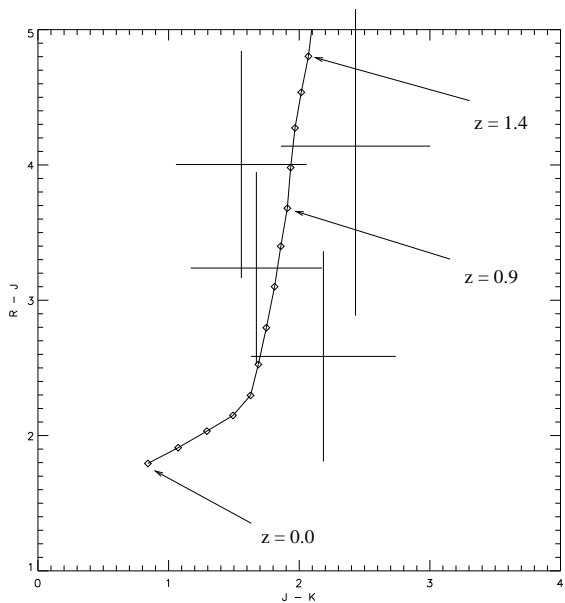,width=\linewidth,clip=}
\caption{The host galaxies of the radiosources in the $RJK$
colour-colour plane. The radiogalaxies are shown as large crosses, with the size of the cross
indicting the photometric erros. The line shows the colours of a non-evolving E0 galaxy as it is
redshifted from $z = 0.0$ (lower left), with symbols marking intervals of $\Delta z = 0.1$. The
centroid of the radiosources in colour-colour space is at  $z \approx 1$.}
\label{fig:rjk_colcol_plot}
\end{center}
\end{figure}

The optical and infrared magnitudes of the radiogalaxies immediately imply that they lie at
significant redshift. In particular, given the $K$-band Hubble diagram for faint radiosources (such
as the 7C sample of Willott, Rawlings and Blundell, 2001), the $K \approx 18$ magnitudes of the
TOC~J0233.3+3021 radiogalaxies suggest that they are $L^*$ galaxies at redshifts of $z\approx 1.0$.
We next consider the $RJK$ colours of the radiogalaxies. It is now well-established that faint
radiogalaxies at $z \sim 1$ are hosted by Extremely Red Objects (EROs) with colours as red as $R-K
\approx 6$.  (e.g. Dunlop et al. 1996; Willott, Rawlings and Blundell, 2001). This is consistent
with a picure in which faint radiosources at these redshifts are hosted by old elliptical galaxies.
Comparing the the radiogalaxies with galaxies in the $z = 1.27$ cluster ClG~J0848+4453 found by
Stanford et al. (1997), we find a close similarity in $RJK$ colour-magnitude diagrams
(Fig. \ref{fig:rjk_colmag_plot}).
If we plot the $RJK$ colours of the
radiogalaxies against the track of a redshifted E0 galaxy, we find
that they cluster about the $z \approx 1$ region on colour-colour space
(Fig.  \ref{fig:rjk_colcol_plot}).
Lastly, the faint red continuum detected in the optical spectra of the
radiogalaxies is also consistent with their lying at $z \approx 1$. 

These optical and infrared data do not alone demonstrate unequivocally
that this overdensity of radiogalaxies is a real cluster. However, the
coincidence between the overdensity of sources and the CMB decrement
leads us to conclude that we are seeing the SZ effect of a massive
cluster, and that most, if not all, the radiogalaxies are cluster
members.

\subsection{Cluster properties inferred from the SZ effect}

From the SZ flux on the shortest spacing, we can now estimate the gas
mass of the cluster. We used {\sc Profile} (Grainge et al. 2001b) to
model the cluster as a spherical King-profile gas distribution, with a
gas temperature of 7keV, $\beta$-parameter 0.65, and find the best fit
value of the central electron density ($n_0$) to the data for varying
core radius $\theta_C$.  While the temperature and $\beta$ parameter
are unknown in the absence of deep X-ray imaging, the values we assume
are typical of those which have been been measured in X-ray selected $
z > 0.5$ clusters (see, e.g., Grego et al. 2001 for a summary). For an
$\Omega_M = 0.3$, $\Omega_{\Lambda} = 0.7$ cosmology, with $H_0 =
65$kms$^{-1}$Mpc$^{-1}$, we find that the gas mass within a radius of
1~Mpc is proportional to $\theta_C^{1/2}$. We place the constraint
$\theta_C>20^{\prime\prime}$ from our measurement that the flux on the
second shortest RT baseline is less than 200~$\mu$Jy. This core radius
corresponds to a physical distance of 178~kpc and the appropriate best
fit model has $n_0 = 10^4~\rm m^{-3}$ and a central decrement of
$900~\mu$K.  We therefore estimate that the minimum gas mass of the
cluster is $5 \times 10^{13} M_{\odot}$. Assuming that the total mass
of the cluster is ten times greater than this, TOC~J0233.3+3021 is
clearly a massive cluster similar to those identified at $z > 0.5$ in
X-ray surveys such as EMSS.

The accurate measurement of the properties of this population of
high-$z$ massive clusters is a significant goal in observational
cosmology. In this Letter, we have demonstrated that current SZ
observations are effective in confirming high-$z$ cluster candidates
which have been selected by other means; SZ investigations of other
TOC candidates are continuing. The new generation of SZ survey
telescopes will not be restricted to the use of `signposts' to
high-$z$ clusters, and promise to detect the majority of the distant
cluster population.

\section{Conclusions}

We have observed the field of TOC~J0233.3+3021, a cluster of
milliJansky radiosources selected from the NRAO VLA Sky Survey (NVSS),
with the Ryle Telescope (RT) at 15 GHz, and in the optical and
near-infrared with the McDonald 2.7-m, Calar Alto 3.5-m, and
UKIRT. Our conclusions are as follows:

\begin{enumerate}

\item After subtraction of the baseline-independent fluxes of the
radiosources, we find a seven-sigma decrement on the RT shortest
baseline of $-695$ $\mu$Jy.

\item The radiosource host
galaxies have magnitudes of $K \approx 18$, $J \approx 20$, $R
\approx 24$. From the infrared magnitude-redshift relation for
radiosources, the galaxies must lie at $z \gta 0.5$. The similar $RJK$ colours strongly suggest that
these galaxies are evolved ellipticals at $z \sim$, and the galaxy magnitudes are consistent
with the $R - J$ colour-magnitude diagrams of known $z \gta 1$ clusters.

\item We therefore conclude that we have identified a strong SZ
decrement with a cluster of galaxies at $z \sim 1$. To produce
the SZ decrement, the cluster must have a mass-temperature product
similar to that of massive Abell clusters at low redshift; we estimate
a minumum gas mass of $5 \times 10^{13} M_{\odot}$. These observations
support the growing evidence for a population of massive clusters at
$z \gta 1$.

\item These observations demonstrate  the
effectiveness of combining searches for AGN `signposts' to clusters
with the redshift-independence of the SZ effect.  Investigations of
this and other high-$z$ cluster candidates selected from NVSS are
continuing.

\end{enumerate}

\section*{ACKNOWLEDGEMENTS}

We thank staff of the the Cavendish Astrophysics group for the
operation of the Ryle Telescope, which is funded by PPARC. We thank
Joe Tufts for help with the $R$-band observing, and Chris Willott for
help with the ISIS spectroscopy. GC acknowledges a PPARC Postdoctoral
Research Fellowship. HJB and SDC acknowledge PPARC PhD
studentships. RD acknowledges support from a British Marshall
Scholarship. UKIRT is operated by the Joint Astronomy Centre on behalf
of PPARC. The WHT is operated on the island of La Palma by the Isaac
Newton Group in the Spanish Observatorio del Roque de los Muchachos of
the Instituto de Astrofisica de Canarias. This material is based in
part upon work supported by the Texas Advanced Research Program under
Grant No. 009658-0710-1999


\section*{REFERENCES}

Blanton, E.L., Gregg, M.D., Helfland, D.J., Becker, R.H., 
White, R.L., 2000, ApJ, 531, 118\\
Chapman, S.C., McCarthy, P.J., Persson, S.E., 2000, AJ 120, 1612\\
Condon, J. J., Cotton, W. D., Greisen, E. W., Yin, Q. F., Perley, R. A.,
Taylor, G. B., Broderick, J. J., 1998, AJ, 115, 1693\\
Dunlop, J.,  Peacock, J.,  Spinrad, H.,  Dey, A.,  Jimenez, R.,
Stern, D., Windhorst, R., 1996, Nature, 381, 581\\
Fabian, A.C., Crawford, C.S., Ettori, S., Sanders, J.S., 2001, MNRAS
in press, astro-ph/0101478\\
Grainge, K., Jones, M., Pooley, G., Saunders, R., Baker, J., Haynes, T., 
Edge, A., 1996, MNRAS, 278, L17\\
Grainge, K.,  Grainger, W.F., Jones, M.E., Kneissl, R., Pooley, G.G., 
Saunders, R., 2001a, MNRAS submitted, astro-ph/0102496\\
Grainge K., Jones M. E., Saunders R., Pooley G. G., Edge A.,
Kneissl R., 2001b, MNRAS submitted\\
Grainger, W.F., Das, R., Grainge, K., Jones, M.E., Kneissl, R., Pooley, G.G.,
Saunders, R., 2001, MNRAS submitted, astro-ph/0102489\\
Grego, L., et al., 2001, ApJ submitted, astro-ph/0012067\\
Holder, G.P., Mohr, J.J., Carlstrom, J.E., Evrard, A.E., Leitch, E.M., 2000, ApJ, 544, 629\\
Jenkins, A., et al, 2001, MNRAS, 321, 372\\
Joy, M., et al, 2001, ApJL, submitted, astro-ph/0012052\\
Kneissl, R., Jones, M. E., Saunders, R., Eke, V.R., Lasenby, A.N.,
Grainge, K., Cotter, G., 2001, MNRAS submitted\\
Rosati P., Stanford S.~A., Eisenhardt P.~R., Elston R., Spinrad H., Stern D., Dey A., 1999, 
AJ,  118, 76\\ 
Sunyaev, R. A., Zel'dovich, Ya. B., 1972. Comm. Astrophys. Sp. Phys., 4, 173.\\
Stanford, S.A., Elston, R., Eisenhardt, P.R.,  Spinrad, H., Stern, D.,
Dey, A., 1997, AJ, 114 2232\\
Willott, C.J., Rawlings, S., Blundell, K.M., 2001. MNRAS in press,
astro-ph/0011082\\

\end{document}